\newif\ifarxiv
\arxivtrue

\ifarxiv
\documentclass[acmsmall,nonacm,screen]{acmart}
\else
\documentclass[acmsmall,screen]{acmart}
\fi
\usepackage[utf8]{inputenc}
\usepackage{xcolor,soul}
\usepackage[noline,noend]{algorithm2e}
\usepackage[frozencache,cachedir=minted-cache]{minted}
\usepackage[nounderscore]{syntax}
\usepackage{wrapfig}
\usepackage{pgfplots}
\usepackage{pgfplotstable}
\usetikzlibrary{pgfplots.groupplots}

\pgfplotsset{compat=1.17}

\SetKwProg{CRDT}{crdt}{}{}
\SetKwInput{InitialState}{initialState}
\SetKwProg{Function}{function}{}{}
\SetKwProg{Merge}{merge}{}{}
\SetKwProg{Operation}{operation}{}{}
\SetKwProg{Query}{query}{}{}
\SetKwBlock{Loop}{loop}{end}
\SetKw{Break}{break}

\newcommand{\system}{Katara}

\title{\system: Synthesizing CRDTs with Verified Lifting}

\author{Shadaj Laddad}
\orcid{0000-0002-6658-6548}
\affiliation{
  \institution{University of California, Berkeley}
  \country{USA}
}
\email{shadaj@cs.berkeley.edu}

\author{Conor Power}
\orcid{0000-0002-0660-5110}
\affiliation{
  \institution{University of California, Berkeley}
  \country{USA}
}
\email{conorpower@cs.berkeley.edu}

\author{Mae Milano}
\orcid{0000-0003-3126-7771}
\affiliation{
  \institution{University of California, Berkeley$^{\dag}$}
  \country{USA}
}
\email{mpmilano@cs.berkeley.edu}
\thanks{$^{\dag}$ also at Sutter Hill Ventures}

\author{Alvin Cheung}
\orcid{0000-0001-6261-6263}
\affiliation{
  \institution{University of California, Berkeley}
  \country{USA}
}
\email{akcheung@cs.berkeley.edu}

\author{Joseph M. Hellerstein}
\orcid{0000-0002-7712-4306}
\affiliation{
  \institution{University of California, Berkeley$^{\dag}$}
  \country{USA}
}
\email{hellerstein@cs.berkeley.edu}

\newif\ifcomments
\commentstrue
\ifcomments
    \providecommand{\shadaj}[1]{{\protect\color{purple}{\bf [shadaj: #1]}}}
    \providecommand{\conor}[1]{{\protect\color{red}{\bf [conor: #1]}}}
    \providecommand{\alvin}[1]{{\protect\color{purple}{\bf [alvin: #1]}}}
    \providecommand{\mae}[1]{{\protect\color{blue}{\bf [mae: #1]}}}
    \providecommand{\joe}[1]{{\protect\color{teal}{\bf [joe: #1]}}}
    \providecommand{\jmh}[1]{{\protect\color{teal}{\bf [joe: #1]}}}
    \providecommand{\david}[1]{{\protect\color{green}{\bf [david: #1]}}}
    \providecommand{\davidmwei}[1]{{\protect\color{pink}{\bf [david wei: #1]}}}
    \providecommand{\kaushik}[1]{{\protect\color{orange}{\bf [kaushik: #1]}}}
    \providecommand{\justin}[1]{{\protect\color{green}{\bf [justin: #1]}}}
    \providecommand{\mingwei}[1]{{\protect\color{rhodamine}{\bf [mingwei: #1]}}}
    \providecommand{\rithvik}[1]{{\protect\color{red}{\bf [rithvik: #1]}}}
    \providecommand{\nc}[1]{{\protect\color{pink}{\bf [nc: #1]}}}
     \providecommand{\accheng}[1]{{\protect\color{olive}{\bf [accheng: #1]}}}
  \else
    \providecommand{\shadaj}[1]{}
    \providecommand{\conor}[1]{}
    \providecommand{\alvin}[1]{}
    \providecommand{\mae}[1]{}
    \providecommand{\joe}[1]{}
    \providecommand{\jmh}[1]{}
    \providecommand{\david}[1]{}
    \providecommand{\davidmwei}[1]{}
    \providecommand{\kaushik}[1]{}
    \providecommand{\justin}[1]{}
    \providecommand{\mingwei}[1]{}
    \providecommand{\rithvik}[1]{}
    \providecommand{\nc}[1]{}
    \providecommand{\accheng}[1]{}
\fi

\begin{abstract}
Conflict-free replicated data types (CRDTs) are a promising tool for designing scalable, coordination-free distributed systems. However, constructing correct CRDTs is difficult, posing a challenge for even seasoned developers. As a result, CRDT development is still largely the domain of academics, with new designs often awaiting peer review and a manual proof of correctness. In this paper, we present {\system}, a program synthesis-based system that takes sequential data type implementations and automatically synthesizes verified CRDT designs from them. Key to this process is a new formal definition of CRDT correctness that combines a reference sequential type with a lightweight ordering constraint that resolves conflicts between non-commutative operations. Our process follows the tradition of work in verified lifting, including an encoding of correctness into SMT logic using synthesized inductive invariants and hand-crafted grammars for the CRDT state and runtime. {\system} is able to automatically synthesize CRDTs for a wide variety of scenarios, from reproducing classic CRDTs to synthesizing novel designs based on specifications in existing literature. Crucially, our synthesized CRDTs are fully, automatically verified, eliminating entire classes of common errors and reducing the process of producing a new CRDT from a painstaking paper proof of correctness to a lightweight specification.
\end{abstract}

\ifarxiv
\setcopyright{cc}
\setcctype[4.0]{by}
\else
\setcopyright{rightsretained}
\acmPrice{}
\acmDOI{10.1145/3563336}
\acmYear{2022}
\copyrightyear{2022}
\acmSubmissionID{oopslab22main-p454-p}
\acmJournal{PACMPL}
\acmVolume{6}
\acmNumber{OOPSLA2}
\acmArticle{173}
\acmMonth{10}
\fi



\begin{document}

\maketitle

\section{Introduction}
In today's interconnected world, there is an ever-growing need to
write correct, scalable distributed programs that can serve users at
any location with low latency. Many such applications rely on {\it
 distributed state}, which in turn is often {\it replicated} at
multiple locations. Replication addresses many common concerns in
distributed systems: it can lower latency by keeping a copy of data
close to each client, improve availability by increasing the
odds that some replica is on a reachable machine, and enhance
the scalability of request handling by allowing the overall load of
requests to be partitioned across replicas.

However, programmers are trained to write sequential programs and often struggle to write correct distributed programs that make concurrent updates to replicated state. As programs execute, nodes may update their replicas at different
times or in different orders, which can cause replicated state to diverge and often results in erroneous application behavior. Our broad goal, first articulated in~\cite{hydro}, is to get the best of both worlds: allow developers to \emph{write familiar sequential code} and use \emph{program synthesis to lift} that code to an efficient distributed implementation.

A classic solution to bridge the gap between sequential and distributed semantics is to introduce \textbf{coordination}. Coordination allows all replicas to agree upon the order of execution; as a result, each replica can delay the application of early-arriving operations, which, if applied eagerly, would lead to divergence. Protocols for coordination (e.g. Paxos \cite{paxos}, Raft \cite{raft}, and Zookeeper \cite{zookeeper}), offer a general-purpose solution to preserving sequential execution. However such
approaches are prohibitively expensive for many applications---particularly at global scale~\cite{cassandra,tardis,keepingCalm}.
Such coordination weakens the benefits of
replication, as executing operations issued to single replicas now
involves high-latency communication with other nodes. Thus, \emph{avoiding} coordination has become a popular
approach in the design of modern distributed systems \cite{dynamo,hydro}.

One of the most widely-adopted approaches for designing coordination-free
programs is the use of \textbf{conflict-free replicated data types (CRDTs)}
\cite{CRDTs}.  Rather than relying on coordination to decide the order in which operations execute, CRDTs instead
choose to limit their operations to only those which {\em commute}; if
all replicas of a CRDT see the same set of commutative operations, then that CRDT
will always {\em converge} to the same state, eliminating the threat of
permanent replica divergence.  As a result, the system will {\em eventually} reach a {\em
  consistent} state at each replica.  For applications that can accommodate
this {\em eventual consistency} \cite{eventual-consistency}, CRDTs
enable coordination-free state replication. CRDT
interpretations of traditional sequential datatypes, including
shopping carts, maps, sets, and logs, have found wide adoption in both
academia \cite{logoot,jsonCrdt} and industry \cite{leagueOfLegends,riak,akka}.

CRDTs provide a framework for coordination-free replication, but it is left to developers to design individual CRDTs that capture \emph{application-specific} data models. For many, this is out of reach as ensuring convergence and semantic correctness is challenging even for experts~\cite{kleppmannBug}. Furthermore, specifying CRDTs is difficult, as the corresponding operations on sequential types are often {\it not} commutative.  For example, one
may wish to replicate a set that supports both insertions and removals, in which the order of insertions and removals
is critical to the final state.  The desire to
use such data types has given rise to a class of CRDTs that {\it
  almost} match the behavior of a sequential data type.
For example, remove operations will
``appear'' to evaluate before concurrent add operations in the Add-Wins Set CRDT, regardless of the order
in which those operations arrive at each replica. These semantic differences make such CRDTs hard to verify \cite{crdtVerif}. Moreover, these CRDTs require complex logic to capture the effects of operations while ensuring that replicas ultimately converge.

In this paper, we introduce {\system}\footnote{\url{https://github.com/hydro-project/katara}}, an open-source system that automates the process of CRDT creation by leveraging
\textbf{verified lifting}~\cite{verifiedlifting, qbs, dexter}. Using program synthesis techniques, we \emph{lift} annotated implementations of sequential data types in languages like C/C++ to full implementations of nearly equivalent CRDTs. The sequential data types being lifted appear as standard data structures in traditional
software, can include constructs such as branching and loops,
and do not need to come equipped with custom convergence properties.
Users need only to annotate their sequential data types with
a simple function that defines the order in which conflicting operations should \emph{appear} to occur. For example, when lifting a set data structure, a user may choose to order removal
operations before all concurrent addition operations, specifying the
semantics of an Add-Wins Set.
We then {\em automatically verify} synthesized CRDT candidates against a combination of the sequential semantics and user-specified conflict resolution policy.  Crucially, such policies
are easy to specify, requiring only a handful of lines in all our
examples. By automating the process of verifying a candidate design, we can generate complete
implementations of the CRDT's state, operations, and queries without any user intervention.

CRDT designs can be split into two categories: op-based CRDTs and state-based CRDTs. 
In this paper, we synthesize \textbf{state-based CRDTs (CvRDTs)}, as these can be deployed in more environments and can always be translated to op-based CRDTs if necessary \cite{shapiro2011comprehensive}.
State-based CRDTs are defined by a datatype representing their state, and an associative, commutative, and idempotent (ACI) \emph{merge function}, which determines how the states of replicas are combined to reach convergence.
The state type and merge function together form a
join semilattice, with the merge function serving as the join. 

Early CRDT work involves complex state structures, but it was subsequently observed that CRDTs can be assembled via composition of simple join-semilattices on sets, integers, or Booleans~\cite{bloomL,anna}. We leverage this approach in the context of synthesis.
By limiting the search space of state types to compositions of
join-semilattices, we are able to achieve convergence---normally the
most difficult property of CRDTs to verify---{\em entirely by construction}.

With this intuition, we introduce new algorithms
for searching the space of possible CRDT implementations, including
the state representation of the CRDT, operations defined on
it, and the queries by which its state may be observed.
This includes the design of grammars for runtime logic that we search with a Syntax-Guided Synthesis engine \cite{sygus} and a parallelized enumerative search over compositions of semilattices for the state structure used within the CRDT. We apply multiple SMT
encodings of our correctness conditions to quickly prune the CRDT search space and perform unbounded verification. {\system} is able to automatically generate a variety of practical, provably correct CRDT designs for a wide range of specifications.

To summarize, we make the following contributions:

\begin{itemize}
\item We define a CRDT's correctness in terms of its
  operations and queries, and demonstrate how users can specify CRDTs by augmenting a sequential data type with lightweight ordering
  constraints that resolve conflicts between non-commutative operations
  (Section~\ref{specification}).

\item We introduce an SMT encoding of our correctness conditions that
  enables automated verification of CRDTs against sequential data
  types with ordering constraints, along with a bounded variant that enables efficient pruning of the program search
  space (Section \ref{verification}).

\item We design a synthesis algorithm that efficiently searches
  semilattice compositions for the internal state of the CRDT, creates
  grammars for runtime components that guarantee convergence, and
  applies syntax-guided synthesis to generate both the core logic and
  invariants that prove correctness over unbounded executions (Section
  \ref{synth}).

\item We describe a practical implementation of {\system}, including the details of how we automatically generate
  verification conditions from sequential data type implementations in C/C++ and optimize
  performance by synthesizing several candidate CRDTs in parallel
  (Section~\ref{implementation}).

\item We demonstrate how {\system} can automatically lift
  sequential data types into practical CRDTs, and generate alternative
  designs with behavior equivalent to human-designed CRDTs in existing literature (Section \ref{eval}).
\end{itemize}

\section{Motivating Example} \label{example}
The distributed shopping carts problem, made popular by Amazon's Dynamo~\cite{dynamo}, is an essential business problem with a clever coordination-free solution. In the original formulation of this problem, the authors track a mapping of items to non-negative counts representing how many of that item are in the cart. Users can interact with the shopping cart by requesting insertions and removals of items. The cart can also be queried to determine the count of each item during a checkout procedure. The goal is to replicate a single shopping cart across many distributed nodes to improve fault tolerance, while ensuring eventual consistency so that the accumulated states on any node can be used to query the complete cart.

Let us focus on a simplified version of this problem, where each item can only be in the cart at most once---effectively simplifying the cart to a set of items. A developer without a distributed systems background could attempt to implement this as a replicated data type by having the insertions, removals, and queries all operate on a standard hash-set. However, if we deploy this in a distributed setting, we will immediately begin to see issues.

\begin{figure}[h]
    \centering
    \includegraphics[width=\textwidth]{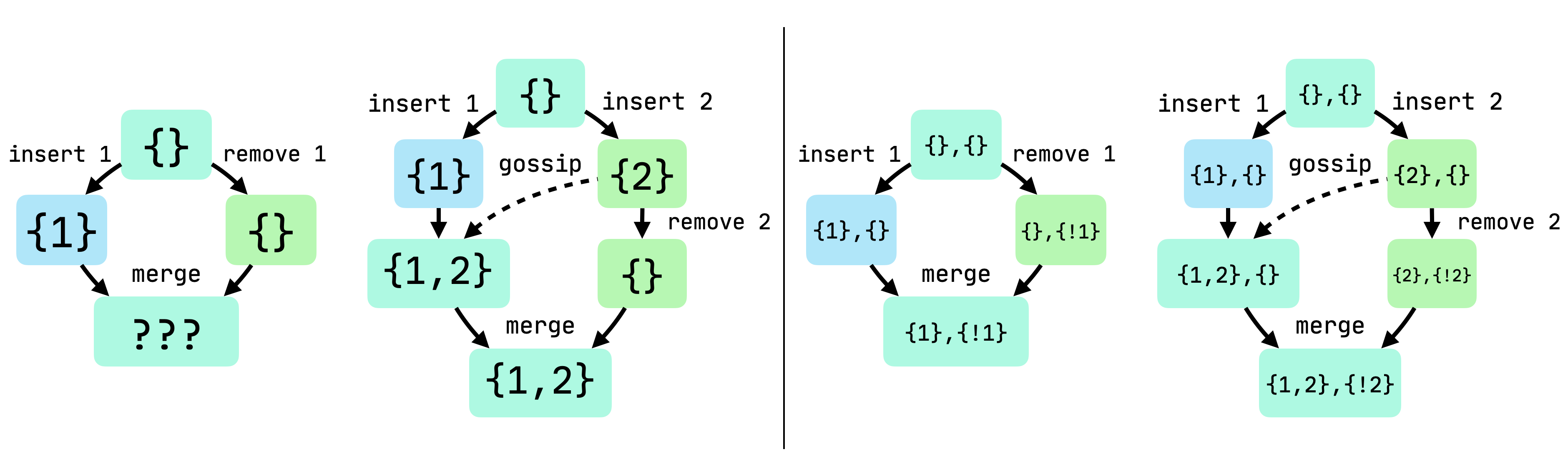}
    \caption{Situations where a sequential type (left) has consistency issues that are resolved by a CRDT (right).}
    \label{fig:broken_crdt_examples}
\end{figure}

Consider the leftmost scenario in Figure~\ref{fig:broken_crdt_examples}, where a shopping cart is replicated across two nodes. When a user sends operations to add and remove the same item, these requests are distributed between the nodes. In this case, the node that receives the insert operation will add the item to its local set, but the node that receives the remove will treat it as a no-op because its local set is currently empty. We may wish to merge the state at these nodes via set union, but this would fail to preserve the locally-ineffective remove operation; it is unclear if this matches user expectations. 

If we periodically share state via gossip \cite{gossip}, new situations emerge where consistency is disrupted. In the execution graph with gossip on the left of Figure~\ref{fig:broken_crdt_examples}, we see an execution where one node processes an insert, and the other processes an insert and then remove of the same element. Even if we merge via set union, we still end up with non-deterministic results depending on when gossip takes place. If the state of the right node is gossiped to the left after the insert but before the remove, the left node will merge the new value into its local state. Even after the remove is processed, merging the two states together will still result in the element 2 being present. But if the gossip were not to take place, the merged state would only have the element 1.

Situations like these have severe correctness implications, and it can be challenging to reason about the many ways a distributed system can break sequential assumptions built into a data type. Furthermore, it is challenging for developers to fix such correctness issues, because doing so involves reasoning about all possible interleavings of operations and their possible effects. Our work aims to tackle this issue by automatically synthesizing a CRDT, which satisfies the property of \emph{convergence} and can be safely replicated in a distributed cluster without requiring coordination. Let's explore how a user would synthesize a shopping cart CRDT with {\system}.

Our synthesis algorithm takes two inputs: a sequential data type that defines the semantics of operations and queries the CRDT will support, and an ordering constraint that specifies how to resolve conflicts between non-commutative operations. We already have the first, since the user has implemented a sequential, single-node shopping cart. For the second, the non-commutative operations in the sequential type are inserts and removes, so the developer may decide that they want the CRDT to resolve conflicting operations by having the removes ``win.'' This can be encoded as a simple pairwise ordering constraint ($opOrder$) that is passed into the synthesis algorithm.

Once {\system} is given this specification, it searches potential state types and runtime logic that are both behaviorally correct and convergent. Along the way, our synthesis algorithm prunes out candidates by using bounded verification to quickly check correctness against short sequences of operations. Eventually, the synthesis engine will produce a provably correct CRDT. For our running example, we might get the CRDT in Figure~\ref{fig:synthesis_example} with an internal state of two sets $(s1, s2)$.

\begin{figure}[h]
    \centering
    \hspace{0.02\textwidth}
    \begin{minipage}[t]{.5\textwidth}
      \centering
      user input
      \begin{minted}[fontsize=\small]{c}
set* init_state() { return set_create(); }

set* next_state(set* state, int add, int v) {
  if (add == 1) return set_insert(state, v);
  else return set_remove(state, v);
}

int query(set* state, int v) {
  return set_contains(state, v);
}
      \end{minted}
      
      \vspace{0.1cm}

      $opOrder(o_1, o_2)$: $o_{1,add} = 1 \lor o_{2,add} \neq 1$
    \end{minipage}
    \begin{minipage}[t]{.42\textwidth}
      \centering
      synthesized design
      \begin{algorithm}[H]
        \CRDT{ShoppingCart}{
        \InitialState{$(\{\}, \{\})$}
        \Merge{$(a_1,a_2),(b_1,b_2)$}{
          \Return{$(a_1 \cup b_1, a_2 \cup b_2)$}
        }
        \Operation{$(s, add, value)$}{
          \Return{$merge(s,$}
          \eIf{$add = 1$}{
            $(\{value\}, \{\})$
          }{
            $(\{\}, \{value\})$
          }
          $)$
        }
        \Query{$((s_1, s_2), value)$}{
          \Return{$value \in (s_1 \setminus s_2)$}
        }
        }
      \end{algorithm}
    \end{minipage}
    \caption{A user-provided sequential reference and the CRDT design synthesized by {\system}.}
    \label{fig:synthesis_example}
\end{figure}

Readers who are familiar with literature on CRDTs may recognize the implementation above as a Two-Phase Set \cite{shapiro2011comprehensive}, one of the classic CRDTs that mimics the behavior of a set while guaranteeing convergence in distributed execution. This synthesized implementation is provably correct for the given sequential specification with the operation reordering, so we can deploy it in our distributed application without having to worry about manually proving complex CRDT properties. We can see on the right of Figure~\ref{fig:broken_crdt_examples} that our CRDT now consistently handles the execution graphs that had conflicts with the naive implementation. Without any baked-in knowledge of existing CRDTs, {\system} is able to automatically generate such implementations that were previously only designed by distributed systems researchers, and can even generate new undiscovered designs by searching the wider space of CRDT structures.

\section{Specifying CRDTs with Sequential Data Types} \label{specification}
In verified lifting, a key piece of the puzzle is specifying correctness of the synthesized program in terms of the reference code. Past work has been focused on transpiling legacy functions into high-level DSLs, a domain in which correctness has a relatively simple definition: the synthesized logic must produce the same output as the original code for any valid input. 
When lifting CRDTs, however, our challenge is different: we are not just finding a function that produces the correct input/output pairs, but rather finding a \emph{stateful} data structure that produces the correct outputs for \emph{unbounded sequences} of method invocations---including mutations and queries. In this section, we develop a formal model for sequential data types and CRDTs and introduce the correctness conditions for a CRDT to match a sequential specification.

We model sequential data types as a combination of two functions: a state transition ($st(s, o)$) that takes in the current state and an operation (a tuple of client-provided parameters) and returns an updated state, and a query function ($query(s, q)$) that takes the current state and a query (similarly, a tuple of client-provided parameters) and returns some data of any type. Sequential types also define an initial state ($initialState$) that is updated as operations are processed. This model can handle a wide range of sequential data types, including those that do not separate updates from queries, since independent operation/query endpoints can be combined into a single $st$/$query$ function and we do not restrict the logic inside those functions.

On the CRDT side, we have a similar interface with one additional function to handle gossip from distributed nodes. In our discussion, we will distinguish the functions corresponding to a CRDT candidate from the sequential data type by attaching an asterisk to the CRDT names. Because the CRDT uses a different state type than the sequential structure, we also mark CRDT states with an asterisk ($s^{*}$). Just like before, we have an initial state ($initialState^{*}$), state transition ($st^{*}(s^{*}, o))$ returning a value of the CRDT state type, and query ($query^{*}(s^{*}, q)$) which together model how the CRDT processes requests over time. In addition, we introduce a merge function ($merge^{*}(s_1^{*}, s_2^{*})$), which is used to merge a node's local state with gossip received from other nodes so that the replicated data converges.


CRDTs have two key components to their correctness: they must respond appropriately to operations and queries, and they must converge under eventual consistency. We guarantee the convergence by construction with grammar restrictions on the state (Section~\ref{synth_structure}) and runtime logic (Section~\ref{synth_runtime}). In this section, we focus on specifying correct operations and queries by comparing the behavior of the CRDT to a reference sequential data type.

\subsection{Specifying CRDTs with Operation Sequences}
Because the sequential data type and the CRDT may use different internal state representations, we cannot directly compare instances of them by comparing their states. As a result, our definition of correctness must reason about the \emph{user-observable behavior} of the data type over \emph{unbounded sequences of interactions}. Both the sequential type and the CRDT have two components a user can interact with: the state transition and the query. Since queries are the only way for users to observe data, we want to ensure that after processing any sequence of user interactions, the sequential type and the CRDT respond identically to any query. Because queries do not modify the state, we can simplify this condition: a sequential type and CRDT are equivalent if both return the same result to an arbitrary query after processing an arbitrary sequence of operations.

But we have to go a step further to justify this correctness definition, since CRDTs can be executed in a distributed system. When we replicate the data type, operations will non-deterministically arrive at nodes in different orders. In addition, the use of gossip protocols in the cluster results in additional state updates when a node merges its local state with a state received from another node. As a result, instead of having a totally-ordered sequence of operations taking the initial state to the final one, we instead have a partial order---a directed acyclic graph---that captures the many different orderings of operations that could be seen by different replicas.

Thankfully, operations on a CRDT that satisfies an eventual consistency policy are commutative: the CRDT state depends only on the \emph{unordered set} of observed operations. Even for CRDTs with causally consistent semantics, operations can be made commutative by extracting sources of causality such as timestamps into operation parameters, making them commutative \emph{modulo} the causality~\cite{cav19crdtverification}. Therefore, as long as the provided CRDT is convergent and has commutative operations, we can reduce verification of a distributed CRDT execution to the sequential case via an arbitrary flattening of the execution graph.

\subsection{Resolving Commutativity with Operation Orderings}
So far, our correctness conditions require strict equivalence of the sequential data type and the synthesized CRDT. However, the additional requirement of operator commutativity on the CRDT means that many sequential types with non-commutative operations, including common ones like sets and maps, cannot directly correspond to an equivalent CRDT but instead are mapped to many popular CRDT variants that make different semantic compromises.

In {\system}, users can define these semantic adjustments through \textbf{operation orderings}, which loosen the correctness requirements to only verify sequences of operations following a specific ordering constraint. This approach, reminiscent of past work on CRDT specification~\cite{burckhardt2014}, minimizes user effort (by leveraging verified lifting) and enables automated verification (Section~\ref{verification}). Formally, a user can define a partial order $\mathit{opOrder}(o_1, o_2)$, which returns true when a call to $o_2$ is allowed to occur after a call to $o_1$.

As an example of the effect of this ordering, recall the shopping cart we lifted in the motivating example. In our sequential data type, inserts and removals do not commute, so no CRDT exists that strictly matches its semantics. However, we can resolve the conflict by introducing an ordering between the operations. If we specify that removes take place before inserts, we get a specification of a Grow-Only Set, which treats removes as no-ops. On the other hand, if we specify that removes take place after inserts, we get a specification of a Two-Phase Set, where elements can be inserted, then removed, but not inserted again.

Instead of having to consider all potential interleavings of operations in a distributed system, operation orderings make it possible for users to specify the distributed behavior in terms of a \emph{sequential execution model}. This makes it possible for non-experts to use our synthesis algorithm, since they can reason about the operation ordering with the existing sequential data type.
This general approach of ordering operations to resolve commutativity allows us to synthesize a wide variety of CRDTs by applying different orderings to simple sequential data types. Intuitively, ordering the non-commutative operations of the sequential type transforms the semantics to be effectively commutative, since any ordering of non-commutative operations will always be reordered into the same sequence by the $\mathit{opOrder}$ constraint. Since the CRDTs we are verifying are already guaranteed to have commutative operations, it then suffices to verify correctness with sequences of operations that follow this order.

\subsection{Operation Orderings with Time} \label{sec:ordering_timestamps}
A popular pattern when designing replicated data types is to place ``distributed timestamps'' on all operations, which makes it possible to introduce sequential semantics without losing convergence. For example, the Last-Writer-Wins Set is a classic CRDT that timestamps its operations; each operation supersedes any conflicting operations that have strictly earlier timestamps. Because distributed timestamps are only a partial order, there can be conflicts among operations that are incomparable in time, which are handled with remove- or add-wins semantics.

When a user provides a sequential data type and ordering specification, we allow them to enable a flag to introduce timestamps to each operation. This flag augments every operation with an integer timestamp $o_t$, which is computed by the local node at runtime using a source that can be mapped to an integer value. We use Lamport timestamps \cite{lamportTimestamps} as this source, which guarantees that causally ordered events will have accordingly ordered integer timestamps. We then augment the $\mathit{opOrder}$ to order operations first by their timestamp, and then apply the user-defined ordering on operations with the same timestamp: $$\mathit{opOrder}(o_1, o_2) \triangleq (o_{1,t} < o_{2,t}) \lor ((o_{1,t} = o_{2,t}) \land \mathit{opOrder}_{\mathit{orig}}(o_1, o_2))$$


When we introduce time, we also must introduce constraints on the operations we consider in our correctness conditions to avoid degenerate cases with illegal timestamps. We do this through an additional user-defined function $\mathit{opPrecondition}(o)$, which checks that an operation is valid. When timestamps are enabled, we define the precondition as $\mathit{opPrecondition}(o) = o_t > 0$ to ensure the operations we check have valid timestamps. Users can also add constraints to this precondition based on domain knowledge, such as if an operation parameter will always be positive.

\section{Automated CRDT Verification} \label{verification}
Now, we must encode this formal definition of CRDT correctness to enable the automated verification of candidate CRDT designs. We tackle this by encoding correctness in SMT logic, which allows us to use solvers like Z3 \cite{z3} and CVC5 \cite{cvc5} to automatically prove correctness or find counterexamples. However, these solvers cannot directly reason about unbounded sequences of operations, so we must break down the correctness conditions into an inductive proof that reasons about individual state transitions.

\subsection{State Equivalence} \label{sec:verification_equivalence}
First, we focus on checking CRDT correctness without considering the ordering constraint. To build this inductive proof, we need a way to reason about the relationship between the states of the sequential data type and candidate CRDT after processing the same, arbitrary sequence of operations. To do this, we choose to relate the states of the CRDT and sequential data type implementations in the style of a bisimulation.

\newcommand{\equivalent}{\ensuremath{\textit{equivalent}^{*}}}
\newcommand{\initialState}{\ensuremath{\textit{initialState}}}
\newcommand{\opPrecondition}{\ensuremath{\textit{opPrecondition}}}
\newcommand{\opOrder}{\ensuremath{\textit{opOrder}}}
\newcommand{\query}{\ensuremath{\textit{query}}}
\newcommand{\orderWithState}{\ensuremath{\textit{orderWithState}^{*}}}

\newcommand{\relation}{\ensuremath{\textit{relation}^{*}}}
\newcommand{\invariant}{\ensuremath{\textit{invariant}^{*}}}

\begin{figure}[h]
  \begin{align*}
    (1)&~ \textit{\equivalent}(\initialState, \initialState^{*})\\
    (2)&~ \forall{s,s^{*},o}: (\equivalent(s, s^{*}) \land \opPrecondition(o)) \implies \equivalent(st(s, o), st^{*}(s^{*}, o))\\
    (3)&~ \forall{s,s^{*},q}: \equivalent(s, s^{*}) \implies \query(s, q) = \query^{*}(s^{*}, q)
    \end{align*}
    \caption{The verification rules that constrain CRDT synthesis to preserve the source semantics}
    \label{fig:verif_rules}
\end{figure}

We encode this proof by introducing the \textbf{state equivalence} function, which relates the states of the sequential reference and synthesized CRDT. Formally, if the reference and synthesized data types are in equivalent states, then after both process an arbitrary sequence of operations they will return the same result to any query. Intuitively, the equivalence function describes which states of the sequential data type correspond to states of the CRDT that capture the same queryable knowledge. Furthermore, the equivalence function captures invariants about the CRDT that filter unreachable states from the verification conditions. With this function available as the inductive invariant of our bisimulation, we can now encode our correctness conditions in SMT logic.

We begin our conditions in Figure \ref{fig:verif_rules} with rule (1), that the initial states of the sequential data type and CRDT must be equivalent. Next, we build the inductive proof that carries equivalence all the way to the final query. We start by encoding the query constraint in rule (3), where we query the reference and synthesized implementations in equivalent states. For this condition, the query results are of the same type so we can directly check for equality. Intuitively, this rule requires equivalence to be a guarantee that queries on the two states return the same result. However, since equivalence deals with not only the current state but also queries on the future states, the equivalence condition may need to be stronger. The condition that forces this strengthening is the inductive step of our proof in rule (2), which checks that if the two data types are in equivalent states, then after executing the same operation they should still be in equivalent states.

\subsection{Enforcing Operation Orders with Invariant Synthesis}
So far, our verification conditions ignore the presence of the user-defined operation ordering (\opOrder), which specifies how the CRDT should handle conflicting non-commutative operations. In {\system}, we implement two encodings of the constraints imposed by this ordering: one that supports unbounded verification but requires synthesizing additional invariants, and one that is only suitable for bounded verification but enables efficient exploration of the program space. In our end-to-end algorithm (Section~\ref{sec:synth_e2e}), we use the bounded encoding first to quickly prune out candidate state structures.

To introduce the ordering constraint to the unbounded verification conditions, we take the approach of strengthening the inductive hypothesis by synthesizing additional invariants. The key insight in this approach is that CRDT states are accumulated by merging updates produced by each operation, so we can enforce orderings that use simple comparisons (such as equality or greater/less than) on the \emph{accumulated state} instead of the individual operations in the history of the CRDT. For example, in the shopping cart scenario where inserts are ordered before removes, we know that an insert is in-order when the set of removed elements is empty.

Formally, we introduce another synthesized Boolean function $\orderWithState(s^{*}, o)$, which returns true if the operation $o$ satisfies the ordering constraint against the history of operations \emph{implied} by the state of the CRDT. This function must be true when executing any operation in a correctly ordered sequence, which in turn ensures that the CRDT correctly handles all executions that satisfy the ordering constraint. By enforcing the user-provided ordering in terms of just the CRDT state, for which we already have inductive invariants, we are able to completely avoid the issue of separately reasoning about the history of operations that have been applied to the CRDT; we directly prove correctness in the unbounded case.

\begin{figure}[h]
  \begin{align*}
    (1)~& \forall{o}: \equivalent(\initialState, \initialState^{*}) \land\\
    &(\opPrecondition(o) \implies \boldsymbol{\mathit{orderWithState}^{*}}\mathbf{(\boldsymbol{\mathit{initialState}^{*}}, o)})\\
    (2)~&\forall{s,s^{*},o_1,o_2}: (\equivalent(s, s^{*}) \land \opPrecondition(o_1) \land \boldsymbol{\mathit{orderWithState}^{*}}\mathbf{(s^{*},o_1)})\\
    &\implies (\equivalent(st(s,o_1), st^{*}(s^{*},o_1)) \land ((opOrder(o_1, o_2) \land opPrecondition(o_2)) \implies\\
    &\boldsymbol{\mathit{orderWithState}^{*}}\mathbf{(st^{*}(s^{*},o_1), o_2)}))\\
    (3)~&\forall{s,s^{*},q}: \equivalent(s, s^{*}) \implies \query(s, q) = \query^{*}(s^{*}, q)
  \end{align*}

    \caption{The verification rules updated to use a synthesized invariant for ordering constraints}
    \label{fig:verif_rules_invariant}
\end{figure}

To start, we update the base case to require that any operation executed in the initial state must be in-order. Then, we update the inductive step to enforce the correctness of $\orderWithState$ on \emph{all} operations in an ordered sequence, by extending rule (2) to reason about pairs of adjacent operations. We introduce a precondition that checks if the first operation being executed is in-order with the state, and enforce the transitive property that $o_2$ be in-order with the state after $o_1$ is processed if it is pairwise in-order after $o_1$. When these conditions are satisfied, we have a proof that our CRDT matches the sequential data type under the operation ordering for any unbounded execution.

\subsection{Optimizing Verification with Quantified Queries}
Because it is critical to verifying a CRDT candidate, the $\equivalent$ invariant must be synthesized alongside the other runtime logic. But with equivalence defined in terms of just the reference and synthesized states, synthesis can quickly become infeasible when the internal states involve large, unbounded structures such as sets and maps (Section~\ref{synth_structure}), which would require the synthesizer to generate higher-order logic like reductions to compare the states. But we can significantly reduce this burden by noticing that beyond the inductive step, equivalence is only used as a precondition for checking that the sequential data type and CRDT return the same response for a \emph{specific query}. Therefore, we can reduce the synthesis burden by introducing an additional parameter $q$ to {\equivalent} so that it is only responsible for checking that the states are observationally equivalent \emph{for a given query}.

\begin{figure}[h]
    \begin{align*}
    (1)~& \forall{o,\mathbf{q}}: \equivalent(\initialState, \initialState^{*}, \mathbf{q}) \land\\
    &(\opPrecondition(o) \implies \mathit{orderWithState}^{*}(\initialState^{*}, o))\\
    (2)~&\forall{s,s^{*},o_1,o_2, \mathbf{q}}: (\equivalent(s, s^{*}, \mathbf{q}) \land \opPrecondition(o_1) \land \mathit{orderWithState}^{*}(s^{*},o_1))\\
    &\implies (\equivalent(st(s,o_1), st^{*}(s^{*},o_1), \mathbf{q}) \land ((opOrder(o_1, o_2) \land opPrecondition(o_2)) \implies\\
    &\mathit{orderWithState}^{*}(st^{*}(s^{*},o_1), o_2)))\\
    (3)~&\forall{s,s^{*},q}: \equivalent(s, s^{*}, \mathbf{q}) \implies \query(s, q) = \query^{*}(s^{*}, q)
    \end{align*}
    \caption{The verification conditions with the additional query parameter for equivalence}
    \label{fig:rules_with_q}
\end{figure}

By giving the equivalence function a specific query, the synthesized logic can now focus on comparing the parts of each state that are relevant to that query. For example, when synthesizing a CRDT that uses maps, this can result in significant simplifications like only checking one key. We update the verification conditions by adding a new quantifier for the query to rules (1) and (2). In rule (3), we simply pass the existing query variable to the {\equivalent} function. We define these updated conditions in Figure~\ref{fig:rules_with_q}. An additional optimization this enables in Section~\ref{sec:synth_core} is to move the postcondition of rule (3) into the structure of {\equivalent}, which further reduces the burden on the synthesizer since it would otherwise have to discover this condition by itself.

\subsection{Solution Pruning with Bounded Operation Logs}
The unbounded verification conditions, while necessary to prove correctness for the CRDT, have a large performance impact on synthesis since they require both the CRDT and $\mathit{orderWithState}^{*}$ to be synthesized simultaneously. To reduce this impact, we employ a two-phase synthesis approach where we synthesize the core logic of candidate CRDTs with verification conditions that check a bounded number of operations (and therefore do not require additional invariants), and separately synthesize the invariants to prove unbounded correctness.

There is one key modification to the base verification rules that we need to make: the state transition should only be checked when the operation being processed is in-order according the user-provided function. To encode this, we add an additional variable to the verification conditions ($\sigma$) that stores a bounded log of operations that have been processed. Because this operation log has a statically known bound, we can lower it to a fixed set of variables corresponding to each element of the list and avoid having to involve more complex theories. Note that the log cannot be used by any of the synthesized logic, since its only role is to aid verification.

\begin{figure}[h]
    \begin{align*}
    \text{list in-order/valid}: lio(\sigma) \triangleq &\text{ }\forall{i}: (i < |\sigma|) \implies (opPrecondition(\sigma[i])~\land\\
    &((i < |\sigma| - 1) \implies opOrder(\sigma[i], \sigma[i + 1])))\\
    \text{list coherent}: lc(s^{*}, \sigma) \triangleq &\text{ }s^{*} = fold(\sigma, initialState^{*}, st^{*})
    \end{align*}
  \begin{align*}
    (1)~& \forall{q}: \equivalent(\initialState, \initialState^{*}, q)\\
    (2)~& \forall{s,s^{*},\boldsymbol{\sigma},o,q}: (\equivalent(s, s^{*},q) \land \opPrecondition(o) \land\\
    &\boldsymbol{\mathit{lio}}\boldsymbol{(\sigma)} \land \boldsymbol{\mathit{lc}}\boldsymbol{(s^{*}, \sigma)} \land \boldsymbol{\mathit{opOrder}}\boldsymbol{(\sigma[|\sigma|-1], o)})\\
    &\implies \equivalent(st(s, o), st^{*}(s^{*}, o),q)\\
    (3)~& \forall{s,s^{*},q}: \equivalent(s, s^{*}, q) \implies \query(s, q) = \query^{*}(s^{*}, q)
    \end{align*}

    \caption{The verification rules updated to use operation logs for bounded ordering constraints}
    \label{fig:verif_rules_history_list}
\end{figure}

Then, we update the state transition verification rule to add a precondition that the operation log is in-order and coherent with the state of the CRDT. First, we check that every pair of operations in the log are in-order, since the operation log is a quantified variable in the SMT encoding and may have out of order values. Then, we verify that the synthesized state equals the result of folding over the log with the synthesized state transition. Since the operation log is bounded, this collapses into a bounded number of state transitions and can be efficiently verified by an SMT solver. Finally, we add a condition that the operation being applied in rule (2) is in-order with the existing log, which can be checked by comparing it against the last operation (since the ordering constraint is transitive).

Note that we do not need to introduce the ordering and coherence preconditions to the query verification conditions in rule (3) even though that rule operates on arbitrary input states. Because we still synthesize the $\equivalent$ function to relate the CRDT and sequential data type, we do not need to constrain \emph{how} we reach the synthesized state being evaluated, just that any instance of the sequential reference deemed equivalent will return the same response to the query. It is left to the synthesizer to introduce any invariants necessary to avoid checking queries on unreachable states.
Together, these rule modifications are summarized in Figure \ref{fig:verif_rules_history_list}. With bounded operation logs, this encoding enables efficient synthesis of the state transition and query functions.

\section{CRDT Synthesis Algorithm} \label{synth}
Now that we have a formal specification of correctness that can be verified by an SMT solver, we are ready to define the synthesis algorithm for CRDT implementations. Our end-to-end algorithm requires only two pieces of user input: the sequential data type written in a standard imperative language and the operation ordering that resolves conflicts. Our synthesis algorithm optionally takes a set of Boolean flags that enable synthesis of advanced designs, such as those that use timestamps (discussed in Section~\ref{sec:ordering_timestamps}) or have non-idempotent operations (which we explore later in this section).

There are four core components to synthesize: the type of the internal state, the initial state, the state transition function, and the query function. We must also synthesize the $equivalent$ and $orderWithState$ invariants from the previous section to enable verification. As discussed before, the synthesized CRDT may use a completely different state structure than the source, which adds a new layer of complexity because the choice of state type affects the search space for each synthesized function. Therefore, our synthesis algorithm uses multiple phases to generate candidates of runtime logic for a range of potential state types.

In Section~\ref{specification}, we explained that our verification conditions only check the user-observable behavior of the CRDT, but do not verify that the CRDT implementation meets the convergence properties. Instead of checking these properties through verification conditions \cite{cav19crdtverification}, we craft our CRDTs in a way that satisfies these properties \emph{by construction}. Inspired by past work on designing coordination-free distributed systems \cite{bloomL,anna}, we synthesize CRDTs that use \textbf{semilattice compositions} for their internal state, which makes it straightforward to enforce monotonicity and commutativity since these are properties of the semilattice join.

\subsection{State Synthesis} \label{synth_structure}
To explore candidate state structures for the CRDT, we use the classic synthesis approach of defining a grammar and iteratively processing deeper structures. Because we focus on compositions of semilattices, our grammar consists of simple rules for primitives, sets, maps, and tuples.

For primitive types, we include semilattice definitions based on Booleans and integers, which are sufficient to lift a wide variety of sequential data types. For Booleans, we have the \texttt{OrBool} lattice, which is a Boolean that has $\bot = false$ and is merged with $\lor$. For integers, we provide the \texttt{MaxInt} semilattice, which merges integers by taking the maximum. Beyond the primitives, we include a semilattice definition for \texttt{Set<T>}, which can have a non-lattice type \texttt{T} for elements; the only constraint on \texttt{T} is that it supports equality.

Our lattice definitions for composite data structures are more complex. First, we offer the \texttt{LexicalProduct<A, B>} semilattice, where A and B are themselves semilattices. In this semilattice, the first element has priority over the second when determining the ordering of two instances. This type is especially useful for CRDTs that use timestamps to have recent operations override older ones, but need to perform a merge over the underlying values when the effects of concurrent operations are combined. We define the lattice join for \texttt{LexicalProduct} as:

$$(a_1, b_1) \sqcup (a_2, b_2) = \begin{cases} 
  (a_1, b_1) & a_1 > a_2 \\
  (a_2, b_2) & a_2 > a_1 \\
  (a_1 \sqcup a_2, b_1 \sqcup b_2) & \text{otherwise}
\end{cases}$$

This definition respects the lattice axioms of associativity, commutativity, and idempotence. Furthermore, these tuples can be nested to form tuples of arbitrary arity. We also support the \texttt{FreeTuple<A,B>} lattice, which simply joins elements pairwise (i.e. $(a_1, b_1) \sqcup (a_2, b_2) = (a_1 \sqcup a_2, b_1 \sqcup b_2)$) and can similarly be nested to form tuples of arbitrary arity.

In some cases we may not know the desired arity of a \texttt{FreeTuple} in advance, or we may not need all the ``fields'' of such a tuple in a given execution. To address this, we offer a \texttt{Map<K, V>} semilattice, where \texttt{K} can be any type that supports equality, and \texttt{V} is a semilattice. Our maps support common operations such as insertions with the same semantics as regular maps, except when inserting keys that are already present in the map. Instead of overwriting the value, we use the lattice join for the value type to combine the existing value with the one being inserted. This carries over to our definition of the lattice join for maps themselves, where we insert the entries of both maps, with keys that are present in both maps having their values merged according to their join:

$$m_1 \sqcup m_2 = \{k_i: \begin{cases} 
  m_1[i] & (k_i \in m_1) \land (k_i \not\in m_2) \\
  m_2[i] & (k_i \not\in m_1) \land (k_i \in m_2) \\
  m_1[i] \sqcup m_2[i] & (k_i \in m_1) \land (k_i \in m_2)
\end{cases})\}$$

Again, this definition respects the standard lattice axioms. Given these semilattice types, we can construct the grammar in Figure~\ref{fig:lattice_grammar} that defines the space of compositions to explore. Our grammar covers a large space of semantics, since the available types encode core capabilities such as free and lexicographically-ordered semilattice products (via maps and tupling) and general semilattice representations (sets).
In our end-to-end synthesis algorithm, we explore types in this grammar with iteratively increasing depth bounds and attempt to synthesize the runtime component of the CRDT for each one. Note that we only include \texttt{FreeTuple} in the top-level $\mathit{latticeList}$ for CRDTs that need multiple semilattices in their state.

\begin{figure}[h]
    \centering
    \begin{minipage}[t]{.7\textwidth}
    \begin{grammar}
    <latticeList> ::= <latticeType> | FreeTuple(<latticeType>, <latticeList>)

    <latticeType> ::= OrBool | NegBool | MaxInt
    \alt Set(<type>) | Map(<type>, <latticeType>)
    \alt LexicalProduct(<latticeType>, <latticeType>)
    
    <type> ::= Bool | Int
    \end{grammar}
    \end{minipage}

    \caption{The grammar defining compositions of semilattices we explore during synthesis.}
    \label{fig:lattice_grammar}
\end{figure}

\subsection{Runtime Synthesis} \label{synth_runtime}
With our state structure selected, we can now move on to synthesizing the runtime logic. Our algorithm for runtime synthesis proceeds in two phases: a first step that synthesizes the core logic with the bounded operation log verification conditions, and a second that synthesizes the additional invariants required for unbounded verification. By using a two-phase approach, we are able to significantly improve the end-to-end synthesis performance of our algorithm by pruning out state structure candidates for which no runtime implementation satisfies even the bounded conditions. In addition, this approach reduces the number of invariants that must be synthesized simultaneously with the CRDT logic, which further improves efficiency.

\subsubsection{Core Logic Synthesis} \label{sec:synth_core}
We derive significant power from our choice to implement the internal state of the CRDT via a semilattice.  First, we observe that, as lattice join is compositional, we can define the merge function as the lattice join on the internal state; this in turn is derived directly from the state's constituent lattices.  Next, we observe that we can also implement {\em operations} in terms of this lattice join: we define
our state transition as $st^{*}(s^{*}, o) = merge^{*}(s^{*}, f^{*}(o))$, where $f^{*}$, which returns a lattice value of the same type as $s^{*}$, is
the function that we actually synthesize.  This choice grants us monotonicity, commutativity, associativity, and idempotence entirely for free, derived from the lattice join ($merge^{*}$) itself.

Along with the state transition, we synthesize the query
function that is used to pull information out of the CRDT. There are
no convergence restrictions on the query since it does not mutate the state, leaving only the sequential reference as a source of
constraints on its synthesis. As a result, we do not need to craft the
query function in any special way, and can let the synthesis engine
drive the search of the query logic.

To support the inductive step of the verification conditions, we must synthesize the equivalence function. This function has two intuitive roles: (1) a \emph{cross-state relation} that identifies which states of the sequential data type and the CRDT are observationally equivalent, and (2) a \emph{CRDT state invariant} that is needed to strengthen the inductive hypothesis of the correctness proof. Following this intuition, we split the synthesis of the equivalence function into components for each role. As discussed when we introduced the query parameter to $\equivalent$ in Section~\ref{sec:verification_equivalence}, we seed the equivalence function with a check that both states respond identically to the given query. This means that our equivalence function has the form $\equivalent(s, s^{*}, q) \triangleq \query(s, q) = \query^{*}(s^{*}, q) \land \relation(s, s^{*}) \land \invariant(s^{*})$, where $\relation$ and $\invariant$ are synthesized.

Because the embedded query comparison already filters out most states that immediately return different query responses, we can improve synthesis performance with a heuristic that bounds the maximum expression depth of $\relation$ to one less than the other functions. Even with this optimization, we can synthesize complex $\relation$ logic when necessary because the depth is iteratively increased. With the bounded operation log encoding, the $\invariant$ component is unnecessary because we know that the CRDT state can be reached through the explicit log of operations. We will revisit the invariant in Section~\ref{sec:synth_unbounded}, when we synthesize the CRDT using the encoding for unbounded correctness that does not use a log.

\begin{figure}[h]
    \centering
    \begin{minipage}[t]{.45\textwidth}
    $\forall{T, U, O}$
    \begin{grammar}
    <bool> ::= false | true
    \alt <bool> $\land$ <bool> | <bool> $\lor$ <bool>
    \alt $\lnot$ <bool>
    \alt <int> $>$ <int> | <int> $\geq$ <int>
    \alt <T> $=$ <T> 
    \alt<T> $\in$ <Set(T)> | <Set(T)> $\subset$ <Set(T)>
    
    <int> ::= 0 | 1 | <int> $+$ <int> | <int> $-$ <int>
    \alt constants in the sequential source
    \end{grammar}
    \end{minipage}
    \begin{minipage}[t]{.45\textwidth}
    \hfill
    \begin{grammar}
    <Set(T)> ::= $\{\}$ | $\{$<T>$\}$
    \alt <Set(T)> $\cup$ $\{$<T>$\}$ | <Set(T)> $\cup$ <Set(T)>
    \alt <Set(T)> $\setminus$ <Set(T)>
    
    <Map(T, U)> ::= $\{\}$ | $\{$<T>: <U>$\}$
    \alt <Map(T, U)> $\cup$ <Map(T, U)>

    <U> ::= <Map(T, U)>[<T>, default=<U>]
    \alt <Tuple(U, T)>[0] | <Tuple(T, U)>[1]
    \alt \text{input of type }U
    \end{grammar}
    
    if top-level or $U$ is not a Set or Map:
    \begin{grammar}
    <U> ::= \text{if} <bool> \text{then} <U> \text{else} <U>
    \end{grammar}
    \end{minipage}

    \caption{The core grammar used to synthesize the state transition and query functions.}
    \label{fig:grammar_core}
\end{figure}

The synthesized components of the state transition, query, and equivalence functions all use a common core grammar. Similar to past program synthesis work, we generate the grammar in Figure~\ref{fig:grammar_core} based on the type constraints of supported operations and bound it by an iteratively increased depth, discussed further in Section~\ref{sec:synth_e2e}. Our grammar features the core set of operations available on the types we support in our system, such as arithmetic, Boolean logic, and set/map operations. In addition, we include conditionals in our grammar to support branching inside the synthesized logic. Because branches that emit complex types such as sets or maps are expensive to synthesize, we restrict those to the top-level of the expression and seed the condition with any Boolean inputs and equality comparisons for integer inputs.

The astute reader may notice that this grammar does not enforce any of the ACI properties. But this is not a problem! Recall that we are synthesizing a function $f^{*}(o)$ that returns a lattice value to be passed into $merge^{*}$. Therefore, even though some operations in this grammar are not idempotent or commutative, the overall state transition function $st^{*}$ remains associative, commutative and idempotent by construction. The semantics of the operations in our language are largely standard, and we lower the operations directly to the corresponding logic in the SMT solver when possible.

Finally, we synthesize the initial state using a shallow grammar of constructors and relevant constants for each type. We include small integer literals, Boolean constants, and empty instances of sets and maps. In cases in which $\bot$ is defined, the initial state is often synthesized to just be the bottom value of the lattice, but occasionally we want the synthesizer to pick an alternate value to handle queries in the initial state. For example, when synthesizing a Boolean register where concurrent enables are ordered \emph{after} disables, the natural lattice to represent the flag's state is a \texttt{LexicalPair<ClockInt, OrBool>} with $\bot = (0, false)$, but we need the initial state to be $(0, true)$ if the sequential data type starts in a enabled state. By synthesizing this value instead of fixing it to $\bot$, we can synthesize CRDT designs over semilattices that do not define bottom, or where the initial state starts higher in the semilattice order.

\subsubsection{Synthesizing Non-Idempotent Operations}
So far, the state transitions we can synthesize are always idempotent, which is not a requirement of CRDTs in general and prevents us from synthesizing designs such as counters. To resolve this limitation, we use a common trick in replicated distributed systems and relax the idempotence constraint while ensuring that certain state can only have a single writer via constraints on the state transition grammar. We introduce \textbf{node IDs}, which are unique integer identifiers for each node in the cluster that can be used as map keys to separate portions of the state that are tied to each node. With this separation of writable state, we can synthesize non-idempotent operations that update portions of the state that only the current node can write.

Users can introduce node IDs to the synthesis pipeline by enabling a single Boolean flag when the sequential data type has non-idempotent operations. Because synthesizing CRDTs with non-idempotent operators introduces additional variables and a larger grammar, which can impact performance, the flag is disabled by default and must be explicitly enabled by the user based on their knowledge of the sequential data type. In future work, we hope to automate this process by analyzing the sequential reference to automatically detect non-idempotence.

Enabling non-idempotent operators affects two components of the synthesis algorithm: the structure of the synthesized functions and the grammar used for runtime logic. The synthesized component of the state transition, which previously could only read the operation arguments to ensure idempotence and commutativity, is expanded to have access to the CRDT state as well as the current node ID. As a result, we must now synthesize a function with the form $f^{*}(o, s^{*}, \text{currentNodeID})$.

To synthesize CRDT logic that uses node IDs, we add a production rule so that the state transition can read from portions of the state that are owned by the current node, which are the values of maps keyed by a node ID. Similarly, we add a rule that allows the state transition to update portions of the state that the current node owns, by allowing insertions keyed by the current node ID. Finally, we introduce rules to the query grammar for performing reductions over the values of maps keyed by node IDs, which makes it possible to combine the state of each node into a global response to queries. We detail these additional grammar elements in Figure~\ref{fig:grammar_with_nid}.

\begin{figure}[h]
    \centering
    \begin{minipage}[t]{.66\textwidth}
    for the state transition:

    $\forall{V}$
    \begin{grammar}
    <V> ::= <Map(NodeID,V)>[currentNodeID, default=<V>]
    
    <Map(NodeID,V)> ::= <Map(NodeID,V)> $\sqcup$ $\{$currentNodeID: <V>$\}$
    \end{grammar}
    
    \hfill
    
    for queries:
    \begin{grammar}
    <Int> ::= reduce(values(<Map(NodeID,Int)>), $\lambda{a}.\lambda{b}.a + b$, $0$)
    
    <Bool> ::= reduce(values(<Map(NodeID,Bool)>), $\lambda{a}.\lambda{b}.a \lor b$, $false$)
    \alt reduce(values(<Map(NodeID,Bool)>), $\lambda{a}.\lambda{b}.a \land b$, $true$)
    \end{grammar}
    \end{minipage}

    \caption{The additional production rules required to support non-idempotent operations.}
    \label{fig:grammar_with_nid}
\end{figure}

Although these changes to the construction of the state transition may allow it to be non-idempotent (and potentially non-commutative), the synthesized CRDT remains convergent because the only requirement for state-based CRDTs is that the merge function agrees with the state transition. Because our state transition still performs a lattice join with the previous state at the top level, and the non-idempotence/commutativity is restricted to portions of state, each owned by an individual node in the cluster, the merge function remains correct since a node can never receive new information about the portions of state it owns through gossip from other nodes.

\subsubsection{Pruning Grammars with Specialized Types}
While shallow instantiations of these grammars are sufficient to synthesize simple CRDTs, such as grow-only sets, they quickly grow to infeasible sizes when the state structure involves a larger number of nested data structures. Much of the grammar expansion comes from a conflation of types that can have distinct semantic meanings, resulting in production rules like arithmetic and comparison operations being unnecessarily introduced.

\begin{figure}[h]
    \centering
    \begin{minipage}{.5\textwidth}
      \begin{minted}[fontsize=\small]{c}
set* init_state() { return set_create(); }

set* st(set* s, int add, int v) { ... }

int query(set* state, int v) { ... }
      \end{minted}
    \end{minipage}
    \begin{minipage}{.42\textwidth}
      \begin{minted}[fontsize=\small]{python}
stateTypeHint = Set(OpaqueInt())
opArgTypeHint = [EnumInt(), OpaqueInt()]
queryArgTypeHint = [OpaqueInt()]
queryRetTypeHint = EnumInt()
      \end{minted}
    \end{minipage}
    \caption{An example of how a sequential data type is annotated with specialized types.}
    \label{fig:specialized_type_example}
\end{figure}

To resolve this, we introduce \emph{specialized} integer types, which represent distinct interpretations of integer values. In {\system}, we have \texttt{OpaqueInt}, which represents an abstract value that does not support arithmetic, \texttt{ClockInt}, which represents a positive timestamp that only supports comparison operations, and \texttt{EnumInt}, which represents values that only support equality. Users can then annotate the functions in their sequential data types, as shown in Figure~\ref{fig:specialized_type_example}, to mark types in the state and operation/query functions that conform to these specialized alternatives. When timestamps are enabled by the user to define richer operation orderings, we automatically add the necessary \texttt{ClockInt} annotations for those values.

By using distinct types for integer inputs, we can avoid searching expressions that, for example, compare timestamps to opaque values. We define grammar rules for these types in Figure~\ref{fig:grammar_with_specialized_ints}. These types are also added to the state structure grammar, but for brevity we omit the changes here.

\begin{figure}[h]
    \centering
    \begin{minipage}[t]{.8\textwidth}
    \begin{grammar}
    <bool> ::= <opaque> $>$ <opaque> | <opaque> $\geq$ <opaque> | <opaque> $=$ <opaque>
    \alt <clock> $>$ <clock> | <clock> $\geq$ <clock> | <clock> $=$ <clock>
    \alt <enum> $=$ <enum>
    
    <clock> ::= 0

    <enum> ::= 0 | 1 | constants in the sequential source
    \end{grammar}
    
    \hfill
    
    $\forall{T, U}$
    
    \begin{grammar}
    <U> ::= reduce(values(<Map(T,U)>), $\lambda{a}.\lambda{b}.a \sqcup b$, $\bot$)
    \end{grammar}
    \end{minipage}

    \caption{The production rules for specialized integer types and semilattice reductions.}
    \label{fig:grammar_with_specialized_ints}
\end{figure}

With the grammars defined for all three functions, we can apply a syntax-guided synthesis algorithm to explore the space of CRDT implementations and use the bounded operation log verification conditions to automatically verify candidates using an SMT solver. The bounds used in this phase start at very small values but are incrementally increased based on feedback from later phases of the synthesis algorithm, which we discuss in further detail in Section~\ref{sec:synth_e2e}.

\subsubsection{Invariant Synthesis for Unbounded Verification} \label{sec:synth_unbounded}
After the first synthesis phase produces a CRDT design that passes bounded-log verification, we must synthesize additional invariants to check the CRDT against the unbounded verification conditions. We must re-synthesize the equivalence function with the CRDT state invariant included, since the unbounded conditions depend on the invariant to exclude unreachable CRDT states. In addition, we synthesize the $\orderWithState$ function so that the unbounded conditions can reason about operation orderings.

The CRDT state invariant only has access to the CRDT state, which helps reduce the size of the grammar generated. We seed the invariant with an explicit condition that checks if the state is valid according to the relevant semilattice definitions. Each semilattice in our state grammar comes with logic for checking validity, such as that the integer values for clocks are at least zero. By automatically including these checks, we further reduce the burden on the synthesizer to discover properties needed for the inductive proof. The rest of the invariant is synthesized using the same type-based grammar as the other functions. Note that we do not need to synthesize the relation component of equivalence, since that was already synthesized in the bounded-log phase.

Synthesizing $\orderWithState$ is a bit more complex. Since the role of this function is to determine whether an operation is in-order while only having access to the CRDT state, this function often needs to combine information from large portions of the state rather than just manipulating data associated with specific keys. For example, when synthesizing a CRDT that uses clocks to order operations, $\orderWithState$ will need to check that the timestamp of the given operation is greater than all existing timestamps in the state. But it is challenging for syntax-guided synthesis engines to reason about arbitrary reductions, so we must reduce the complexity of the grammar.

We tackle this by noting that reductions (such as collecting the highest timestamp) use the semilattice join of the type being accumulated. This has intuitive backing as well, since we can check if a single value is at least as high in the semilattice order as several others through a single comparison against the semilattice join over those values. Based on this observation, we add a rule in Figure~\ref{fig:grammar_with_specialized_ints} to compute reductions using the lattice join for all relevant lattice values in the state. Note that we support reductions over map values, which are of a type in <$\mathit{latticeType}$>, but not sets because their elements may not be semilattices.

With these additional production rules, we can synthesize $\invariant$ and $\orderWithState$ for the candidate CRDT. With the invariant grammars configured and the existing $st^{*}$ and $\query^{*}$ functions from the previous phase, we return to the synthesis engine with the unbounded verification conditions. At this point, we are verifying all scenarios the CRDT is expected to correctly handle, so if we successfully synthesize the invariants we have a provably correct CRDT design!

\subsection{End-to-End Synthesis Algorithm} \label{sec:synth_e2e}
Now that we have the search space for state structures and runtime logic defined, we can synthesize the entire CRDT from scratch by simultaneously exploring both spaces. In our end-to-end algorithm, we apply multiple logic synthesis phases and verification modes to create provably correct CRDTs while also pruning the program space early in the synthesis algorithm.

\begin{algorithm}[h]
    \Function{search($\mathit{ref}, \mathit{opOrder}$)}{
        \For{$\mathit{depth} \gets (2..\infty)$}{
            \For{$s \gets \mathit{semilatticeCompositions}(\mathit{depth})$}{
                $\mathit{logBound} \gets 2$

                $\mathit{p2Depth} \gets \mathit{depth}$
                
                \Loop{
                    $\mathit{p1Synth} \gets \mathit{synthBoundedLog}(\mathit{ref}, s, \mathit{opOrder}, \mathit{depth}, \mathit{logBound})$
                
                    \If{$\mathit{p1Synth} = \text{unsat}$}{
                        \Break
                    }
                    
                    \BlankLine
                    
                    $\mathit{p2Synth} \gets \mathit{synthUnbounded}(\mathit{ref}, s, \mathit{opOrder}, \mathit{p2Depth}, \mathit{p1Synth})$
                
                    \eIf{$\mathit{p2Synth} = \text{unsat}$}{
                        $\mathit{logBound} \gets \mathit{logBound} + 1$

                        $\mathit{p2Depth} \gets \mathit{p2Depth} + 1$
                    }{
                        \Return{$\mathit{p2Synth}$}
                    }
                }
            }
        }
        \BlankLine
        \BlankLine
    }

    \caption{The end-to-end algorithm for synthesizing a CRDT from scratch.}
    \label{alg:e2e_algo}
\end{algorithm}

At the top level of the algorithm, we iterate over candidate state structures generated from the grammar of semilattice compositions, bounded to the same $\mathit{depth}$ as the runtime logic. For each of these, we then generate the appropriate runtime logic grammars and perform synthesis with the bounded-log verification conditions (with an initial $\mathit{logBound} = 2$). If we fail to synthesize, we can eliminate the candidate state structure from consideration, since there is no synthesizable logic even when the verification is relaxed.

If we successfully synthesize, we can move on to synthesizing the additional invariants for unbounded verification. We combine the synthesized code from the previous phase with the grammars for $\invariant$ and $\orderWithState$, and call out to the synthesis engine again. If we fail to synthesize at this point, it means that either the bounded-log phase returned a buggy implementation or the grammar for invariants was too small. To address this, we return to the bounded-log phase and increment the operation log bound for verification and the grammar depth ($\mathit{p2Depth}$) for invariants. If we successfully synthesize the invariants, we have a provably correct CRDT that we can return to the user. We summarize this process in Algorithm~\ref{alg:e2e_algo}.

\section{Implementation} \label{implementation}
We implement {\system} using an extended version of the framework in Casper~\cite{casper}, which allows us to automatically extract sequential data types written in C and C++ by first compiling them to LLVM and analyzing the IR to generate the equivalent SMT logic. Our implementation also includes wrappers around the Rosette synthesis engine~\cite{rosette} and CVC5 solver~\cite{cvc5}, which we use to perform synthesis and verification.

\subsection{Supported Language Features}
To synthesize a CRDT, {\system} must first extract the semantics of the sequential data type provided by the user. To ensure that the logic implemented by the user can be accurately translated into the SMT logic used for verification, we define a space of programs that can be safely handled. {\system} can handle basic LLVM operations, branches, integer/Boolean primitives, and list/set/map types.

Our analysis can accurately handle primitive types such as integers and Booleans, along with the corresponding arithmetic and logical operations on them. In addition, we provide a set of APIs for lists, sets, and maps that users can build on in their sequential data type. Our analysis automatically recognizes uses of these specialized APIs and lowers them to the corresponding SMT theory. Our framework offers a modular approach to defining the semantics of these types, so it is straightforward to add support for richer data structures such as stacks.

In addition to analyzing the types and operations on them, our framework can extract branches found in the LLVM IR to conditionals in the generated SMT logic. {\system} generates separate specifications of each basic block found in a function, and links them together to define the function as a whole. This approach allows us to handle nested conditionals and early returns without needing additional logic for these cases. In addition to branches, our analysis also handles user-defined functions by inlining them into the top-level function that is lifted.


\ifarxiv
\hfill
\newline
\else
\fi

\subsection{Bounded Data Structure Verification}
When synthesizing with Rosette, we face a limitation that the size of the symbolic state must be a constant, which means that we cannot define verification conditions that operate over unbounded data structures such as lists or sets. To address this limitation, we bound the size of these data structures to a fixed value while performing synthesis. After Rosette returns us a successfully synthesized CRDT, we then pass the result to CVC5, which can perform verification with unbounded data structures when a theory is defined for their behavior.

CVC5 natively supports a theory of sets, and we provide our own set of axioms that define tuples. When maps---which have not yet been modeled in CVC5---are involved in the synthesized CRDT, we fall back to using Rosette for verification with a large bound for the data structures. We hope to improve this in future work by providing a set of complete axioms that enable the solver to reason about unbounded map instances. In the meantime, the bounds we use for the fallback are sufficiently large to consistently produce correct synthesized results.

\subsection{Parallel State Structure Exploration}
Both Rosette, which uses Z3 under the hood, and CVC5 are single-threaded. If we were to naively implement the end-to-end algorithm, we would underutilize the multi-core capabilities of modern systems. But because we control the search of CRDT state structures, and the logic synthesis for each structure candidate is independent of the others, we can drastically speed up the synthesis algorithm by parallelizing across state structures.

{\system} allows users to configure the number of state structures to synthesize logic for in parallel. Based on this parameter, we then instantiate a thread pool and spawn the logic synthesis algorithm for candidate structures on free threads. Our implementation simply returns the first successfully synthesized CRDT from any thread. By exploring more state structures and avoiding situations where synthesis is blocked on a state candidate that is particularly difficult to synthesize logic for, this technique drastically improves the end-to-end synthesis performance.

\section{Evaluation} \label{eval}
In our evaluation, we explore the capability of {\system} to \emph{correctly} and \emph{efficiently} synthesize CRDTs for a variety of sequential data type and operation ordering specifications. We focus on answering the following research questions:

\begin{itemize}
    \item \textbf{RQ1}: Can {\system} produce practical CRDTs based on specifications sourced from literature on coordination-avoidance?
    \item \textbf{RQ2}: What is the effect of pruning structures with bounded-log verification on the overall synthesis performance?
    \item \textbf{RQ3}: Is the grammar of lattice composition sufficiently rich to produce CRDT designs that differ from the canonical implementations in the literature?
\end{itemize}

\subsection{\textbf{RQ1}: Synthesizing Practical CRDTs}
We begin by evaluating the ability of our synthesis algorithm to produce correct CRDTs from scratch for a variety of user-provided specifications. We sourced several sequential data types and operation orderings, summarized in Table~\ref{tab:crdt_specs}, from existing literature on human-designed CRDTs \cite{shapiro2011comprehensive} and coordination-avoidance \cite{ecros}. For each of the benchmarks, we created a minimal implementation of the sequential type in C based on the specifications provided by the source and encoded the operation ordering using the IR provided by the synthesis system. All benchmarks were conducted on a AMD Ryzen 9 3900X processor with 12C/24T and 48 GB of memory, with our implementation configured to use up to 12 threads. We use LLVM 11 to compile and analyze the sequential types, as well as the latest versions of Rosette (4.1) and CVC5 (1.0.2).

\begin{table}[h]
    \centering
    \caption{The set of CRDT specifications used to evaluate our synthesis algorithm.}
    \begin{tabular}{p{3.7cm}|c|c|c|c}
        Benchmark & Source & Specification Size & Timestamps & Non-Idempotent \\
        \hline
        Grow-Only Counter & Shapiro & 21 LoC &  & \checkmark \\
        General Counter & Shapiro & 20 LoC &  & \checkmark \\
        Enable-Wins Flag & De Porre & 21 LoC & \checkmark &  \\
        Disable-Wins Flag & De Porre & 21 LoC & \checkmark &  \\
        Last-Writer-Wins Register & Shapiro & 16 LoC & \checkmark &  \\
        Grow-Only Set & Shapiro & 24 LoC &  &  \\
        Two-Phase Set & Shapiro & 24 LoC &  &  \\
        Add-Wins Set & De Porre & 24 LoC & \checkmark &  \\
        Remove-Wins Set & De Porre & 24 LoC & \checkmark & 
    \end{tabular}
    \label{tab:crdt_specs}
\end{table}

Our overall approach is designed to require minimal user intervention to produce a practical CRDT. As a proxy for this goal, we measured the amount of code required for a user to specify both the sequential data type and the operation ordering that specifies the synthesized CRDT. For all our benchmarks, both of these components can be declared within 25 lines of C (for the data type) and Python (for the operation ordering). The Boolean flags to enable timestamps and non-idempotence are also provided to the system through a Python API. Most of the lines of specification code are for the sequential data type, which a developer using {\system} will likely already have. The operation orderings, which are specific to {\system}, could all be defined in 4 LoC or less of integer comparisons.

\begin{table}[h]
    \centering
    \caption{The performance of synthesizing CRDTs for the benchmark specifications with {\system}.}
    \begin{tabular}{p{2.55cm}|p{2.15cm}|p{2.15cm}|p{5.6cm}}
        Benchmark & Synthesis Time (default) & Synthesis Time (no pruning) & Synthesized State Type \\
        \hline
        Grow-Only Counter & 1m 46s $\pm$ 0.7s & 52s $\pm$ 0.3s & \footnotesize{\texttt{Map<NodeID, MaxInt<Int{>}>}} \\
        \hline
        General Counter & 11m 3s $\pm$ 4s & 13m 21s $\pm$ 4s & \footnotesize{\texttt{FreeTuple<Map<NodeID, MaxInt<Int{>}>,}}

        \footnotesize{\texttt{  Map<NodeID, MaxInt<Int{>}{>}>}} \\
        \hline
        Enable-Wins Flag & 2m 4s $\pm$ 1s & 2m 53s $\pm$ 8s & \footnotesize{\texttt{LexicalProduct<MaxInt<ClockInt>, OrBool>}} \\
        \hline
        Disable-Wins Flag & 1m 46s $\pm$ 3s & 3m 4s $\pm$ 2s & \footnotesize{\texttt{LexicalProduct<MaxInt<ClockInt>, OrBool>}} \\
        \hline
        Last-Writer-Wins Register & 26s $\pm$ 0.2s & 18s $\pm$ 0.8s & \footnotesize{\texttt{LexicalProduct<MaxInt<ClockInt>,}}

        \footnotesize{\texttt{  MaxInt<Opaque{>}>}}\\
        \hline
        Grow-Only Set & 30s $\pm$ 0.2s & 23s $\pm$ 0.1s & \footnotesize{\texttt{Set<Opaque>}} \\
        \hline
        Two-Phase Set & 58s $\pm$ 0.6s & 1m 5s $\pm$ 0.8s & \footnotesize{\texttt{Map<Opaque, OrBool>}} \\
        \hline
        Add-Wins Set & 21m 17s $\pm$ 11s & 1hr 58m $\pm$ 1m & \footnotesize{\texttt{FreeTuple<Map<Opaque, MaxInt<ClockInt{>}>,}}

        \footnotesize{\texttt{  Map<Opaque, MaxInt<ClockInt{>}{>}>}} \\
        \hline
        Remove-Wins Set & 18m 40s $\pm$ 14s & 1hr 52m $\pm$ 1m & \footnotesize{\texttt{FreeTuple<Map<Opaque, MaxInt<ClockInt{>}>,}}

        \footnotesize{\texttt{  Map<Opaque, MaxInt<ClockInt{>}{>}>}}
    \end{tabular}
    \label{tab:synth_times}
\end{table}

When run with our collection of benchmarks, our synthesis algorithm is able to successfully generate designs that conform to all of the specifications, and it identifies the inductive invariants necessary to prove correctness of each CRDT over unbounded executions. We list the average time required to synthesize each CRDT (along with standard deviations) and the state structure of the synthesized result in Table~\ref{tab:synth_times}. Simpler CRDTs, such as the Grow-Only/Two-Phase Set and LWW-Register, can be synthesized by {\system} in a matter of seconds. More complex CRDTs, especially those that use timestamps to order operations such as the Add/Remove-Wins Set, can take on the order of tens of minutes to synthesize. These performance measurements indicate that the composition of multiple synthesis phases allows for many types of CRDTs to be synthesized in a reasonable amount of time.

\begin{figure}[h]
    \centering
    \begin{minipage}{.47\textwidth}
      \begin{algorithm}[H]
        \CRDT{AddWinsSet}{
        \InitialState{$(\{\}, \{\})$}
        \BlankLine
        \Operation{$(s, \mathit{add}, \mathit{value}, \mathit{time})$}{
          \Return{$s\text{ }\sqcup$}
          \eIf{$\mathit{add} = 1$}{
            $(\{\mathit{value}: \mathit{time}\}, \{\})$
          }{
            $(\{\}, \{\mathit{value}: \mathit{time}\})$
          }
        }
        \BlankLine
        \Query{$((s_1, s_2), v)$}{
          $t_1 = s_1[v, \mathit{default}=0]$
          
          $t_2 = s_2[v, \mathit{default}=0]$

          \Return{$t_1 \geq t_2 \land t_1 > 0$}
        }
        }
      \end{algorithm}
    \end{minipage}
    \begin{minipage}{.52\textwidth}
      \begin{algorithm}[H]
        \CRDT{GeneralCounter}{
        \InitialState{$(\{\}, \{\})$}
        \BlankLine
        \Operation{$((s_1, s_2), \mathit{inc}, \mathit{nodeID})$}{
          $\mathit{cur}_1 = s_1[\mathit{nodeID}, \mathit{default}=0]$
          
          $\mathit{cur}_2 = s_2[\mathit{nodeID}, \mathit{default}=0]$

          \Return{$(s_1, s_2)\text{ }\sqcup$}
          \eIf{$\mathit{inc} = 1$}{
            $(\{\mathit{nodeID}: \mathit{cur}_1 + 1\}, \{\})$
          }{
            $(\{\}, \{\mathit{nodeID}: \mathit{cur}_1 + 1\})$
          }
        }
        \BlankLine
        \Query{$((s_1, s_2))$}{
          $r_1 = \mathit{reduce}(\mathit{values}(s_1), \lambda{a}.\lambda{b}.a + b, 0)$
          
          $r_2 = \mathit{reduce}(\mathit{values}(s_2), \lambda{a}.\lambda{b}.a + b, 0)$

          \Return{$r_1 - r_2$}
        }
        }
      \end{algorithm}
    \end{minipage}
    \caption{The Add-Wins Set and General Counter CRDTs our synthesis algorithm is able to produce.}
    \label{fig:set_and_counter_crdts}
\end{figure}

Our synthesis results also show the algorithm discovering a variety of CRDT design techniques without any baked-in knowledge of CRDTs, such as using timestamps to guard data and storing the effects of conflicting operations in separate parts of the state. For example, {\system} synthesizes the CRDT on the left in Figure~\ref{fig:set_and_counter_crdts} for the Add-Wins Set, which supports repeated insertions and removals by using timestamps to have Adds only shadow Removes when they are concurrent. Similarly, on the right side of Fig~\ref{fig:set_and_counter_crdts}, {\system} is able to discover how to use node IDs to handle non-idempotent operations in a counter CRDT, using multiple reductions in the query to take the difference of the accumulated increments and decrements.

Although our work does not focus on the runtime performance of the synthesized code, all of our CRDTs have comparable theoretical performance to human designs in existing literature. In the case of the Two-Phase Set benchmark, our synthesis algorithm comes up with a more efficient state encoding (which we discuss in Section~\ref{sec:alt_crdt}) that simplifies the state to a single integer-to-Boolean map rather than the typical two integer sets. Overall, our synthesis algorithm is able to produce practical, provably correct CRDTs for the variety of specifications in our benchmarks.

\subsection{\textbf{RQ2}: Search Space Pruning}
A key contribution in our runtime synthesis algorithm is the use of two SMT encodings of the correctness conditions: one that can quickly verify CRDT candidates with checks for bounded executions, and one that can prove unbounded correctness but requires the synthesizer to identify additional invariants. In this section, we explore how this two-phase synthesis approach improves the performance of the overall algorithm.

First, we can compare the time that our end-to-end algorithm takes to find CRDTs for each of the benchmarks with and without the pruning optimization. In Table~\ref{tab:synth_times}, we list the synthesis times without the bounded-log phase under the "no pruning" column. Other than lighter benchmarks which synthesize in around minute and have relatively simple CRDT state types, all the benchmarks synthesize faster with the two-phase algorithm. The largest performance improvements come for the CRDTs with the most complex state structures: the General Counter and Add/Remove-Wins Set. In the case of the Sets, we see up to a 5x speedup by using the two-phase approach.

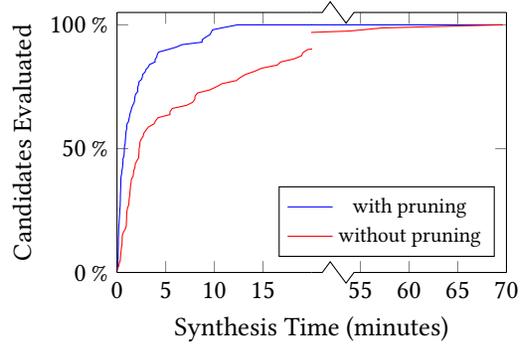
\begin{wrapfigure}{r}{0.5\textwidth}
    \pgfplotstableread[col sep=comma,]{add-wins-bounded-history.csv}\dataBounded
    \pgfplotstableread[col sep=comma,]{add-wins-unbounded.csv}\dataUnbounded
    
    \centering
    \pgfplotsset{
        every non boxed y axis/.style={} 
    }
    \hspace*{-0.2\textwidth}
    \begin{tikzpicture}
    \begin{groupplot}[
        group style={
            group size=2 by 1,
            yticklabels at=edge bottom,
            horizontal sep=0pt
        },
        x filter/.code={\pgfmathparse{#1/60}\pgfmathresult},
        y filter/.code={\pgfmathparse{#1*100}\pgfmathresult},
        height=5cm,
        ymin=0, ymax=105
    ]
    
    \nextgroupplot[xmin=0,xmax=20,
                   xtick={0,5,10,15},
                   axis y line=left,
                   width=0.3\textwidth,
                   xlabel={Synthesis Time (minutes)},
                   x label style={at={(axis description cs:1,-0.3)},anchor=south},
                   ylabel={Candidates Evaluated},
                   yticklabel={\pgfmathprintnumber\tick\,\%}]
    \addplot [blue] table [x={time}, y={percent}]{\dataBounded};
    \addplot [red] table [x={time}, y={percent}]{\dataUnbounded};
    
    \nextgroupplot[xmin=50,xmax=70,
                   xtick={55,60,65,70},
                   axis y line=right, 
                   axis x discontinuity=crunch,
                   width=0.3\textwidth,
                   yticklabels={,,},
                   legend style={at={(0.94,0.06)}, anchor=south east}]
    \addplot [blue] table [x={time}, y={percent}]{\dataBounded};
    \addlegendentry{\footnotesize{with pruning}}
    \addplot [red] table [x={time}, y={percent}]{\dataUnbounded};
    \addlegendentry{\footnotesize{without pruning}}
    \end{groupplot}
    \end{tikzpicture}

    \caption{The distribution of time taken to evaluate candidates for the Add-Wins Set benchmark.}
    \label{fig:bounded_vs_unbounded_plot}
\end{wrapfigure}

For a more nuanced exploration of \emph{why} we see these speedups, we collect the time it takes each synthesis algorithm to either correctly synthesize or prune out each candidate data structure it considers for the Add-Wins Set benchmark. We compare the two algorithms in Figure~\ref{fig:bounded_vs_unbounded_plot}, where we plot a distribution of the percent of candidates (out of 86 total for both) that can be evaluated within a given amount of time. With pruning, all candidate structures can be processed in under 15 minutes, allowing the end-to-end algorithm to quickly reach the state candidate that yields a verified CRDT. Without pruning, many state candidates take up to 20 minutes to be evaluated and there is a long tail of candidates that take up to an hour each. When the CRDT must have a complex state to support the specified semantics, these stragglers have a significant toll on synthesis performance since they block exploration of the program space.

\subsection{\textbf{RQ3}: Alternative CRDT Synthesis} \label{sec:alt_crdt}
Finally, we evaluate the richness of the space of CRDTs our synthesis algorithm explores. In particular, we are interested in the ability of our synthesis algorithm to produce \emph{multiple} CRDT implementations for a \emph{single} specification. Since our combination of a sequential type and operation ordering uniquely defines the user-observable behavior of a CRDT, any alternate designs will be functionally equivalent but may have different memory utilization and performance characteristics.

\begin{wrapfigure}{r}{.42\textwidth}
    \centering
    \begin{minipage}{.42\textwidth}
      \begin{algorithm}[H]
        \CRDT{TwoPhaseSet}{
        \InitialState{$\{\}$}
        \BlankLine
        \Operation{$(m, \mathit{add}, \mathit{value})$}{
          \Return{$m\text{ }\sqcup$}
          \eIf{$\mathit{add} = 1$}{
            $\{\mathit{value}: \mathit{false}\}$
          }{
            $\{\mathit{value}: \mathit{true}\}$
          }
        }
        \BlankLine
        \Query{$(m, v)$}{
          \Return{$\lnot{m[v, \mathit{default}=\mathit{true}]}$}
        }
        }
      \end{algorithm}
    \end{minipage}
    \caption{The novel map-based CRDT that is synthesized for the Two-Phase Set benchmark.}
    \label{fig:map_2p_set}
\end{wrapfigure}

In our exploration of alternate CRDT designs, we focus on the Two-Phase Set, which has moderately complex semantics since its execution has multiple phases: allow inserts, then removes, but not inserts again. When we perform synthesis, the first CRDT that is generated is surprisingly \textbf{not} the 2P-Set from existing CRDT literature, but instead (to our knowledge) a novel design that uses a map to capture the phase of each element. If we continue searching, the algorithm eventually emits the classic 2P-Set, whose state structure is larger.

We list the new design in Figure~\ref{fig:map_2p_set}. There are several clever tricks that the synthesizer comes up with to match the specification while using a simpler state. First, the synthesizer realizes that the behavior of a Two-Phase Set "saturating" after removing an element matches how an \texttt{OrBool} saturates when it becomes true. Next, it discovers that by using a mapping from keys to these values, it can maintain a separate saturating value for each key in the set.

But this still leaves a challenging situation for the initial state. Because we use the saturated $true$ value to represent the element being removed, that means we have to set the value for a key to $false$ when it is inserted the first time. But since $\bot = false$ for an \texttt{OrBool}, that would leave us with no additional state. This is where the final trick is discovered by the synthesizer: to query the map with a default value of $true$. This effectively creates a third state for when the key is not even in the map. By automatically discovering this combination of CRDT design tricks, our synthesis algorithm is able to produce this novel encoding of a Two-Phase Set.

Our synthesis algorithm also produces alternate designs for many other benchmarks, such as using pairs of clocks instead of a \texttt{LexicalProduct} for the enable/disable-wins flag benchmarks. The presence of such alternatives paves the way for future work where we synthesize not only a \emph{correct} CRDT, but a \emph{performant} one according to a cost model that can compare CRDT candidates. Furthermore, the pool of alternative designs may be useful for incrementally re-synthesizing CRDTs as the sequential data type is updated, something we hope to explore in the future.

\section{Related Work}
\subsection{Creating Replicated Objects from Sequential Specifications}

There are a few lines of work that focus narrowly on the same problem
we take aim at here: taking specifications of sequential datatypes and
automatically creating equivalent replicated types. Where these projects differ
from ours is primarily in our use of program synthesis: to our knowledge we
are the first to utilize a search-based synthesis approach to
generate state representations and runtime logic.
Our other differences focus on how we chose to resolve conflicting
operations, and our approach of searching semilattice compositions for the CRDT state.

Gallifrey \cite{gallifrey}, Indigo \cite{indigo}, and ECROs {\cite{ecros}} focus on not just specifying replicated
data types, but also in ensuring that applications which use them do
not see inconsistent state---much as our work verifies correctness
with respect to queries. Beyond the use of program
synthesis in our work, the main point of divergence is in their use of
preconditions and postconditions as a verification tool
to exactly match the behavior of synchronous objects.  Other work also uses this pre/post-condition approach \cite{secros} or shares the goal of matching sequential behavior~\cite{static-local-coordination-avoidance}.
Our goal is not to exactly match sequential behavior; we let
programmers tweak semantics with ordering constraints on conflicting operations.  This allows us to lift all
specified operations into CRDTs, rather than limiting ourselves to operations that already commute (as in Gallifrey) or resorting to explicit synchronization or deferred
re-execution for conflicting events (as in Indigo and ECROs). 

The MRDT line of work \cite{mrdts, certified-mrdts} starts with a similar premise to
ours---creating replicated datatypes from annotated sequential
specifications---but takes a radically different approach.
Our largest differences center around their runtime
system, which is based on a Git-inspired log of versioned
data structures, and in their mode of annotation, which centers around
abstraction and concretization functions. In contrast, our sorting-based annotations are simpler for non-experts to reason about, and our generated CRDTs require only a standard gossip protocol. Additionally, the merge function of MRDTs are generated using a rule-based approach,
whereas our work takes the search-based synthesis approach.

\subsection{Program Synthesis and Verified Lifting}
The synthesis approach taken in this work is directly inspired by
verified lifting \cite{verifiedlifting}, the approach at the heart of work such as Domino \cite{domino}, Casper \cite{casper}, and Dexter \cite{dexter}. With verified lifting, the correctness conditions for synthesizing code in a particular DSL are derived from \emph{existing implementations} in standard languages such as C/C++. Our approach expands on this tradition
primarily by our choice to synthesize entire data
types, instead of just function implementations. This involves more complex
verification conditions that check \emph{behavioral equivalence} between the input
code and the synthesized CRDT, rather than just checking equality of function
outputs, and requires synthesizing more complex invariants to enable verification of unbounded interactions. Past work has explored the formal foundations of specifying CRDT correctness in terms of a sequential reference by layering constraints on how the effects of operations are applied~\cite{burckhardt2014}.
Our introduction of pairwise ordering constraints, guided by this work, enables fully automated verification of CRDTs with lightweight annotations that are easy for non-experts to write.

Few previous systems have attempted to directly apply search-based
program synthesis to the space of replication.  Two
that stand out are Hamsaz~\cite{hamsaz} and Hampa~\cite{hampa}.  Hamsaz uses
programmer-provided invariants to synthesize custom consistency
protocols for the replication of shared data structures.  While its
analysis component is reminiscent of other work, such as Quelea,
and the Indigo line \cite{quelea, indigo, ecros, indigo-ipa}, its novel synthesis
component is of particular interest to this work. Like our work,
Hamsaz uses an SMT encoding of the programmer-specified
semantics to search through potential replication strategies. Hampa \cite{hampa}, a
similar work from the same research group, adds recency to the mix. However, both of these solutions are focused on identifying efficient coordination protocols, rather than \emph{eliminating} coordination altogether. The CRDTs synthesized by our algorithm can be replicated without needing any coordination.

\subsection{Verifying Replicated Data Types}

Many previous systems have also provided verification systems for
manually-implemented CRDTs, checking both convergence properties and
correctness with respect to a specification. Several of these require manual proof effort \cite{replicated-datatype-verification-liquid-haskell, crdtVerif, specifying-crdts}, which make them infeasible for program synthesis approaches that require rapid verification of a large number of candidates without any human involvement.

Of particular note is recent work that explores automated verification of convergence properties via an SMT encoding \cite{cav19crdtverification}, which is especially relevant because we also use SMT to automatically verify CRDT candidates. However, our work differs from this research in \emph{what} is being verified. Because our CRDTs are \emph{convergent by construction}, we do not need to perform any convergence verification. Instead, our verification conditions focus on the \emph{user-observable behavior} of the CRDT, including checking the correctness of queries---something that convergence verification approaches do not include (as queries do not affect convergence).

Other lines of work focus on the general question of correct use of weak consistency \cite{cause-im-strong-enough, ralinearizability}, which is a wider problem that is not specific to CRDTs. Furthermore, these lines of work focus on reasoning about how application invariants can be maintained when using weak consistency under the hood, rather than how different types of state can be replicated in an eventually consistent manner. Indeed, the literature around safely using weak consistency is complementary to our contributions, as they provide a path for developers to safely build applications on top of the CRDTs we synthesize.

There are certain CRDTs that challenge the current limitations of what we can specify in SMT for verification. These make for interesting future research directions. One example is the floating point comparison used in Logoot~\cite{logoot} which would require richer user-specified orderings and extensions to the query language. Another example is the Replicated Growable Array (RGA), which requires tracking sequential data. Our query language does not currently support the types of iterative computations required to reason about sequences, but it has been shown that RGA can be represented as Datalog queries~\cite{kleppmannDatalog} over operations. We see this Datalog representation as a promising direction for enabling synthesis of such advanced CRDTs.

\subsection{Making Replicated Objects Easier to Work With}

Several papers attempt to make the process of programming against weak
consistency tractable. Some do so by exposing
weakly-consistent replicated objects with approachable semantics;
indeed, the original CRDT work fits in this vein \cite{shapiro2011comprehensive}.  Other
such approaches include the CloudTypes work from Microsoft \cite{cloud-types} or work on
allowing an application to safely mix consistency levels via MixT \cite{mixt},
Disciplined Inconsistency \cite{disciplined-inconsistency}, CScript \cite{cscript}, or Red-Blue consistency \cite{redblue}.  Some work automatically chooses consistency levels for the programmer, driving the choice of mixed consistency via program invariants rather than explicit consistency annotations \cite{redblue-automating, quelea, kaki-2018, observable-atomic-consisency-2018, observable-atomic-consisency-2020}.  All work on mixing consistency shares the belief that programmers require stronger consistency for some operations; in contrast, we let users introduce semantic adjustments that eliminate the need for strong consistency.

Other work focuses on ensuring that application executions are
convergent despite the inherent non-determinism of concurrency and
weakly-consistent replication.  These include a long line of work from
Berkeley \cite{bloom,bloomL,alvaro2010dedalus,blazes,conway2014edelweiss}, and the Gallifrey, LASP,
and LVars languages \cite{gallifrey,lasp,lvars}.  While we believe
that whole-language approaches are valuable in this space, 
the CRDT literature typically does not analyze 
programs beyond the boundaries of the CRDT implementation.

\section{Conclusion}
The future of computing is distributed, so it is important to reduce the complexity of developing correct, efficient distributed programs. We believe that verified lifting can be a useful tool towards this goal, by automating much of the process of converting familiar sequential logic into scalable distributed code.
In this paper, we presented a first step in this direction with {\system}, a system that automatically synthesizes CRDT designs from existing sequential data type implementations, requiring only simple annotations that are are easy for developers to reason about.

We formalized the definition of \emph{correctness} for CRDTs in terms of a sequential data type by introducing \emph{operation orderings}, which allow users to define how the CRDT should handle conflicting non-commutative operations. To automate the verification process, we developed SMT encodings of this correctness definition that can be used to check CRDT candidates with a solver. Finally, we explored how compositions of semilattices can be naturally used as the state of a CRDT, and defined grammars for runtime logic and the invariants that enable unbounded verification of correctness. Our end-to-end algorithm efficiently synthesizes CRDTs for a variety of scenarios and produces novel alternatives to human-designed CRDTs. With {\system}, we hope to further unlock the power of distributed systems by making it possible for any developer to automatically replicate their existing data types by synthesizing provably equivalent CRDTs.

\begin{acks}
We thank Audrey Cheng, David Chu, Natacha Crooks, and our anonymous reviewers for their insightful feedback on this paper. 
This work is supported in part by National Science Foundation CISE Expeditions Award CCF-1730628, IIS-1955488, IIS-2027575, DOE award DE-SC0016260, ARO award W911NF2110339, and ONR award N00014-21-1-2724, and by gifts from Amazon Web Services, Ant Group, Ericsson, Futurewei, Google, Intel, Meta, Microsoft, Scotiabank, and VMware. Shadaj Laddad is supported in part by the NSF Graduate Research Fellowship Program under Grant No. DGE 2146752. Any opinions, findings, and conclusions or recommendations expressed in this material are those of the authors and do not necessarily reflect the views of the National Science Foundation.
\end{acks}

\bibliography{bibliography}
\end{document}